\documentclass[reprint,twocolumn,nofootinbib]{revtex4-1}

\usepackage{stackrel}
\usepackage{float}
\usepackage{slashed}
\usepackage{graphicx}
\usepackage{mathrsfs}
\usepackage{amssymb}
\usepackage{amsfonts}
\usepackage{amsmath}
\usepackage{amsthm}
\usepackage{enumitem}
\usepackage{braket}
\usepackage{epstopdf}
\usepackage{dsfont}
\usepackage{setspace}
\usepackage{capt-of}
% ...

\usepackage{color}

\usepackage{fancyhdr}

\nopagebreak
\begin{document}
\title{Two Loop Unification of Non-SUSY SO(10) GUT with TeV Scalars}

\author{T. Daniel Brennan}
\affiliation{NHETC and Department of Physics and Astronomy,
Rutgers University\\
126 Frelinghuysen Rd., Piscataway, NJ 08855}
\email{tdanielbrennan@physics.rutgers.edu}
\begin{abstract}
In this paper we examine gauge coupling unification of the the non-SUSY SO(10) grand unified theory proposed by Babu and Mohapatra  at the two loop level \cite{BM}. This theory breaks down to the standard model in a single step and has the distinguishing feature of TeV non-standard model scalars. This leads to a plethora of interesting new physics at the TeV scale and the discovery of new particles at the LHC. This model gives rise to testable proton decay, neutron-antineutron oscillations, provides a mechanism for baryogenesis, and contains potential dark matter candidates. In this paper, we compute the two loop beta function and show that this model unifies to two loop order around $10^{15}$ GeV. We then compute the proton lifetime, taking into account threshold effects and show that these effects place it above the Super Kamiokande limit \cite{Kamiokande}. 
\end{abstract}

\maketitle

\section{\textbf{Introduction}}

It has been known for quite some time that the standard model, despite being the most predictively successful theory in physics, is far from complete. A flagrant sign of this deficiency came with the observation of neutrino oscillation, indicating that neutrinos have mass in contradiction with the standard model \cite{BMNeutrino,Neutrino1,Neutrino2,Neutrino3,Neutrino4,Neutrino5,Neutrino6,Neutrino7,Neutrino8,Neutrino9,Neutrino10,Neutrino11}. Grand unified theories (GUTs) pose a solution to this problem that is both simple and elegant. GUTs also suggest resolutions to many other issues, including the origin of charge quantization and the relative size of the standard model gauge couplings. In addition, they also suggest resolutions to cosmological problems, such as providing a mechanism to explain baryogenesis and candidates for dark matter.

Among this broad class of theories, the subset of minimal SO(10) theories are arguably the most natural. They describe each generation of the standard model in an irreducible representation of the gauge group and have inherent left-right chiral symmetry.  These theories also provide a natural explanation for the origin of neutrino masses and flavor oscillation through a high scale see-saw mechanism \cite{BMNeutrino,Minkowski,GellMann,Yanagida,Glashow,MohapatraSenjanovic} coming from symmetry breaking due to a \textbf{10}$\oplus$\textbf{126}-plet Higgs sector \cite{book}. 

A key test of grand unified theories is through proton decay experiments. In the standard model, both baryon and lepton numbers are conserved by perturbative interactions and hence protects protons, the lightest baryon, from decay. GUTs however, have additional heavy vector bosons which do lead to proton decay. By nature of gauge symmetry breaking, these heavy gauge bosons have mass on the order of the unification scale which this proton decay mechanism to order $1/M_U^4$, rendering protons effectively stable.

This places a special emphasis on proton decay experiments, making them a powerful tool for determining the viability of GUT models. In experiment, the most commonly sought decay modes are the $p\to e^+\pi^0$ and $p\to \bar{\nu}K^+$. For our purposes, we will focus solely on the first process as it is the dominant decay channel for this class of models assuming that there is small flavor mixing \cite{rev}. By measuring this decay, the latest experiments have placed the lifetime of the proton on the order of $10^{34}$ years \cite{Kamiokande,KamiokandePrelim}. This places a strong constraint on the class of physical theories, providing a minimum requirement for all GUT models. 

However, there are several factors which make the theoretical prediction of the proton decay lifetime hard to estimate. There are two main sources of uncertainty in this calculation. The first source comes from the fact that quark confinement is a non-perturbative effect, so that we must rely on numerical lattice QCD techniques to approximate proton decay amplitudes. The second source of uncertainty, comes from the fact that the low energy effective field theory integrates out many heavy fields which become relevant at higher energies with additional unknown parameters. This uncertainty will be taken into account in calculating the threshold corrections. 

%
%In a recent paper \cite{BM}, Babu and Mohapatra constructed a non-SUSY SO(10) GUT so that there are extra-standard model Higgs bosons which have a mass at the TeV scale. This is interesting because these scalars are potentially accessible to the LHC in run 2 and leads to low energy effects. These TeV sclars lead to predictions of phenomena such as neutron-antineutron oscillation, scalar mediated baryogenesis, and potentially dark matter. These phenomena will be investigated in upcoming papers. This paper aims to qualify this theory as physically plausible by computing the proton lifetime to two loop order and comparing the result to the latest results from the Super Kamiokande experiment. 
%
%In section II of this paper we will review the Babu-Mohapatra model and motivate its study by highlighting some interesting features of the low energy phenomenology. Then we will demonstrate the gauge coupling unification to two loop order and compute the unification scale and mass of the the additional non-standard model scalars. In section IV we use this data to compute the decay lifetime of the proton through the $p\to e^+\pi^0$ channel and show that the predicted value is within reach of the next generation of proton decay experiments. In the final section we propose extensions of the Babu Mohapatra model by including additional Higgs multiplets which both extend the proton lifetime and increase the mass of the light non-standard model scalars. 

\section{\textbf{Babu-Mohapatra Model}}

In a recent paper \cite{BM}, Babu and Mohapatra pointed out that there exists a non-SUSY SO(10) GUT model where the seesaw scale is close to the GUT scale. But the inherent quark-lepton unification in GUTs implies that $\Delta L=2$ breaking required for seesaw give rise to observable $\Delta B=2$ processes leading to neutron-antineutron oscillation. Key to this observation is the existence of a TeV mass color sextet scalar ($\Delta_{u^cd^c}$) transforming like $(6,1,1/3)$ which is necessary for one loop unification. Unification additionally requires the existence of 2 complex weak triplets, which we interpret as arising from an additional ${\bf 45\oplus 45'}$-Higgs multiplets. This collection of scalars lead to one loop unification of the gauge couplings near $10^{15}$ GeV \cite{BM}\footnote{It has also been shown recently in \cite{octet1,octet2} that coupling \\unification can be achieved with other TeV scalar multiplets\\ through more complicated symmetry breaking chains.}. \\
\indent This leads to several experimentally interesting features: (i) the existence of TeV scalars which have the potential to be discovered at the LHC in the near future, (ii) the prediction of proton decay with a short enough lifetime to be observed in the next generation of experiments, (iii) the existence of a color sextet based mechanism for neutron-antineutron oscillation which could be measured in the next generation of experiments \cite{Phillips}, (iv) the existence of a color sextet based mechanism  for GUT scale baryogenesis which is unaffected by electroweak sphaleron processes, and (v) the inclusion of the weak scalar triplets which make an excellent candidate for dark matter due to the $\textbf{45}\oplus\textbf{45}'$ Higgs multiplets' lack of Yukawa coupling to the standard model fields.

Since neutron-antineutron oscillation scales as $M^{-4}_{\Delta_{u^cd^c}}$, it is important to know the precise value of the color sextet scalar mass \cite{BM}. Similarly, to have a reliable prediction for proton decay, one needs a precise value of the unification scale. To accomplish these goals, it is necessary to carry out a two loop analysis of the gauge coupling evolution which can determine the color sextet mass as well as the unification scales more precisely than the one loop analysis of \cite{BM}. It is the goal of this paper to carry out this program.
\section{Coupling Unification}
 The standard method for constructing non-SUSY SO(10) models is by implementing intermediate symmetry groups, each of which are broken at a different energy level, creating a multi-step chain from SO(10) to the standard model. A detailed two loop analysis of many SO(10) breaking chains were carried out in \cite{Chang,Bertolini2,octet2,octet1,octet2}. However the breaking pattern of the model studied in this paper was not included: 

\begin{center}$
SO(10)\stackrel[m_U]{}{\longrightarrow} SU(3)\times SU(2) \times U(1)$\end{center}  %This model, as previously stated, deviates from this practice by assuming that SO(10) is broken directly to the Standard Model at the unification scale:\\

 We will now take a bottom-up perspective and assume that in addition to the standard model, there exists a color sextet $\Delta_{u^c d^c}$ which transforms as (6,1,1/3), and two weak triplets $\omega_i$ which transform as $(1,3,0)$\footnote{Here we use the normalization for the hypercharge so that \\$C_2(R)=\left(\frac{Y}{2}\right)^2$}. The two loop beta function is given by \cite{Machacek,Jones}:

\begin{align}\begin{split}
\beta(g)&=\frac{g^3}{(4\pi)^2}\left\{\frac{2}{3}C_2(F)+\frac{1}{3}C_2(S)-\frac{11}{3}C_2(G)\right\}\\&+\frac{g^5}{(4\pi)^4}\Bigg{\{}\left[2 C_2(F)+\frac{10}{3}C_2(G)\right]S_2(F)\\&-\frac{34}{3}[C_2(G)]^2+\left[4 C_2(S)+\frac{2}{3}C_2(G)\right]S_2(S)\Bigg{\}}
\end{split}\end{align}
\noindent where the different field components are implicitly summed over. Here, $C_2(R)$ and $S_2(R)$ are the Casimir element and Dynkin index of a representation R respectively, and $F, S, G$ correspond to the fermionic, scalar, and adjoint representation of a gauge group $G$. When $G$ is the direct product of semi-simple terms ($G=G_1\times G_2\times...G_n$), the two loop term allows for mixing of the gauge subgroups corresponding to contributions to the gauge boson propagator from field multiplets transforming non trivially under multiple subgroups. These fields add additional terms to the $\beta$ function:
\begin{align}\begin{split}
...+\sum_j\frac{g_i^3 g_j^2}{(4\pi)^4}\left[2 S_2(F_i)C_2(F_j)+4 S_2(S_i)C_2(S_j)\right]
\end{split}\end{align}
where $F_i$ is the sub-representation of $F$ transforming under the gauge group $G_i$. Now, writing the beta function as:
\begin{align}
\beta(g_i)=\frac{g_i^3}{(4\pi)^2}b_i+\sum_k \frac{g_i^3 g_k^2}{(4\pi)^4}b_{ik}
\end{align}
the numerical coefficients are given by:
\begin{align}
b_i=\left(\begin{array}{c}\frac{127}{30}\\ \\-\frac{11}{6}\\ \\-\frac{37}{6}\end{array}\right)\hspace{3mm}b_{ij}=\left(\begin{array}{ccc}
\frac{613}{150}&\frac{27}{10}&\frac{172}{15}\\
&&\\
\frac{9}{10}&\frac{35}{6}&12\\
&&\\
\frac{43}{30}&\frac{9}{2}&\frac{37}{3}
\end{array}\right)
\end{align}
\noindent These match with the 1-loop calculation from \cite{BM}. 

As in \cite{BM}, we will also study the additional case of a second standard model Higgs doublet H(1,2,1/2) which produces the beta function:
\begin{align}
b_i=\left(\begin{array}{c}\frac{13}{3}\\ \\-\frac{5}{3}\\ \\-\frac{37}{6}\end{array}\right)\hspace{3mm}b_{ij}=\left(\begin{array}{ccc}
\frac{64}{15}&\frac{18}{5}&\frac{172}{15}\\
&&\\
\frac{6}{5}&\frac{232}{3}&12\\
&&\\
\frac{43}{30}&\frac{9}{2}&\frac{37}{3}
\end{array}\right)
\end{align}
%
%
%\begin{align}
%\beta_i=\frac{d\alpha_i^{-1}}{dt}=-\frac{b_i}{2\pi}-\frac{1}{8 \pi^2}\sum_{k=1}^3b_{ik}\alpha_k
%\end{align}

Taking into account the mass of the Higgs boson and the threshold corrections \cite{Chang,MohapatraParida,Hall,ParidaPatra}:
\begin{align}
\frac{1}{\alpha_U(M_U)}=\frac{1}{\alpha_i(M_U)}-\frac{\lambda_i(M_U)}{12\pi}
\end{align}
\noindent the $\beta_i$ can be numerically integrated using Mathematica. 
Here we have used:
\begin{align}\begin{split}
&\lambda_i(\mu)=S_2(G_V)-21 S_2(G_V) \log \frac{M_V}{\mu}\\&
+\sum_j S_2(S_j)\log\frac{M_{S_j}}{\mu}+8 \sum_j S_2(F_j) \log\frac{M_{F_j}}{\mu}
\end{split}\end{align}
following the calculation from \cite{Weinberg,Hall} where the $V,S_j,F_j$ are the sets of heavy vector bosons, fermions, and scalar bosons respectively coupled to the gauge group factor $G_i$.

\begingroup
\centering
\includegraphics[width=75 mm,keepaspectratio]{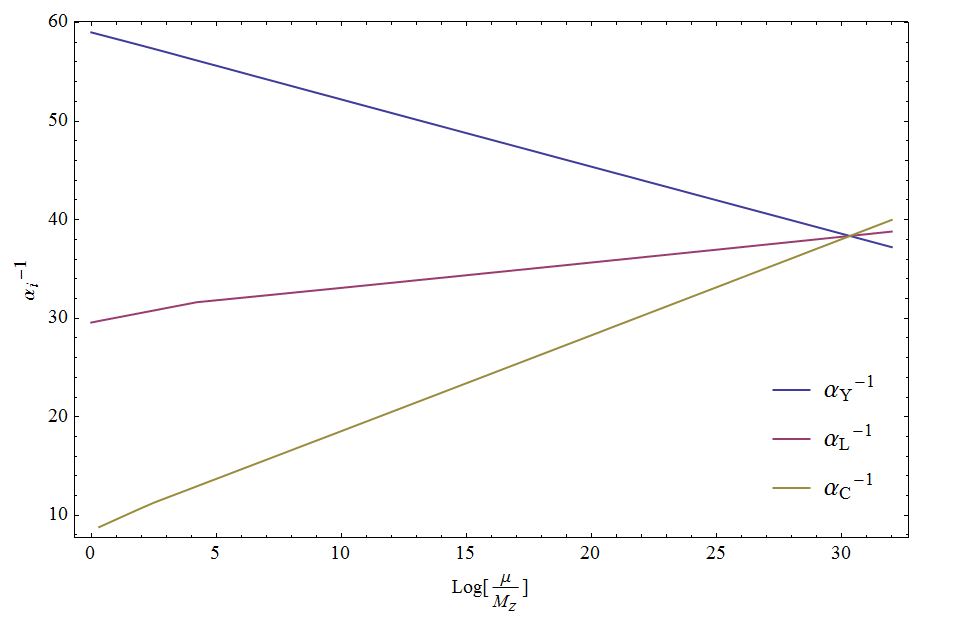}
\captionof{figure}{Running of the standard model gauge couplings with a color sextet $\Delta_{u^c d^c}$ (6,1,1/3) at 1 TeV and 2 weak scalar triplets $\omega_i$ (1,3,0). }%normalized to the mass of the Z boson
\endgroup
\hspace{3mm}

We found that with a single standard model Higgs doublet, the couplings unify at $1.42\times10^{15}$ GeV with $M_\omega=6.16$ TeV and $M_\Delta=1$ TeV, and $\alpha_U(M_U)^{-1}=38.5$. We also found that with two standard model Higgs doublets, the couplings unify at $1.06\times 10^{15}$ with $M_\omega=76.8$ TeV, $M_\Delta=1$ TeV, and $\alpha_U(M_U)^{-1}=38.2$. \footnote{It is important to note that the mass of the scalars are not fixed. Once the mass is fixed for one scalar, the other is fixed by the condition of gauge unification. The given scalar masses were selected to give the maximum proton decay lifetime so that the scalar masses are $\geq 1$ TeV.}
%
%\subsection{Sensitivity}
%
%In general, the unification of this model is not very sensitive to the mass of the scalar multiplets. If use the previous unification parameters for a single standard model Higgs doublet and change the mass of the sextet from 1TeV to 2 TeV we find a very small unification triangle forming (see Figure 2). \\
%
% \begingroup
%\centering
%\includegraphics[width=75 mm,keepaspectratio]{Scaling.png}
%\captionof{figure}{A small unification triangle forms when changing the mass of the color sextet from 1 TeV (solid line) to 2 TeV (dashed line) while keeping the mass of the weak triplets constant.}
%\endgroup
%\hspace{3mm}\\
%
%The size of this triangle is on the order of 110 TeV which is miniscule compared to the triangle formed by the standard model which is on the order of $2.5\times10^5$ TeV.
%

\section{\textbf{Proton Decay}}

From our calculation for the unification scale, we can determine the proton lifetime in this model due to the decay mode: $p\to e^+\pi^0$. In the low energy effective theory, this decay process is dominated by dimension 6 operators. These operators come from integrating out the heavy gauge bosons and heavy Higgs particles. However, from \cite{BM} we know that the standard model constrains the Yukawa coupling to be small enough so that the coupling to the scalar bosons contributions are suppressed relative to that of the heavy gauge bosons by several orders of magnitude. In general there are 5 independent types of such operators which lead to nucleon decay. In the notation of Weinberg, they are denoted \cite{Weinberg,AbbottWise,DM,Cooney}:
\begin{align}\begin{split}
\mathcal{O}^{(1)}_{abcd}&=({d}_{\alpha a R}u_{\beta b R})({q}_{i \gamma c L}l_{j d L})\epsilon_{\alpha \beta \gamma}\epsilon_{ij}\\
\mathcal{O}^{(2)}_{abcd}&=({q}_{i\alpha a L}q_{j \beta b L})({u}_{\gamma c R}l_{d r})\epsilon_{\alpha\beta\gamma}\epsilon_{ij}\\
\mathcal{O}^{(3)}_{abcd}&=({q}_{i\alpha a L}q_{j\beta b L})({q}_{k \gamma c L}l_{l d L})\epsilon_{\alpha \beta\gamma}\epsilon_{ij}\epsilon_{kl}\\
\mathcal{O}^{(4)}_{abcd}&=(q_{i \alpha a L}q_{j\beta b L})(q_{k\gamma c L}l_{l d L})\epsilon_{\alpha\beta \gamma}(\vec{\tau}\epsilon)_{ij}\cdot (\vec{\tau})\epsilon)_{kl}\\
\mathcal{O}^{(5)}_{abcd}&=({d}_{\alpha a R}u_{\beta b R})({u}_{\gamma c R}l_{d R})\epsilon_{\alpha \beta \gamma}
\end{split}\end{align}
\noindent where $\alpha,\beta,\gamma$ denote SU(3) color indices, i,j,k,l are SU(2) isospin indices, a,b,c,d are generation indices, and L and R refer to the chirality. Only 4 of these (1,2,4,5) are relevant for proton decay. Of these only:
\begin{align}\begin{split}
Q^{(1)}&=\mathcal{O}^{(1)}_{1111}=(d_{\alpha R} u_{\beta R})(u_{\gamma L} e_L-d_{\gamma L}\nu^e_L)\epsilon_{\alpha \beta\gamma}\\
Q^{(2)}&=-\frac{1}{2}\mathcal{O}^{(2)}_{1111}=(d_{\alpha L} u_{\beta L})(u_{\gamma R}e_R)\epsilon_{\alpha \beta \gamma}\\
Q^{(3)}&=\tilde{\mathcal{O}}^{(4)}_{1111}=(d_{\alpha L}u_{\beta L})(u_{\gamma L}e_L-d_{\gamma L} \nu^e_L)\epsilon_{\alpha\beta \gamma}\\
Q^{(4)}&=\mathcal{O}^{(5)}_{1111}=(d_{\alpha R}u_{\beta R})(u_{\gamma R} e_R)\epsilon_{\alpha \beta \gamma}
\end{split}\end{align}
lead to the proton decay process: $p\to \pi^0e^+$ \cite{AbbottWise}. Here we used the notation:
\begin{align}
\mathcal{O}_{abcd}^{(4)}=-(\tilde{\mathcal{O}}^{(4)}_{abcd}-\tilde{\mathcal{O}}_{bacd}^{(4)})
\end{align}
We will denote the first part of these operators (the parts with a $e^+$ leptonic term) $\mathcal{O}_{\Gamma \Gamma'}$ where $\Gamma,\Gamma'=L,R$. For example: $\mathcal{O}_{LR}=Q^{(2)}$. In order to calculate the proton lifetime we need to calculate the amplitude:
\begin{align}\begin{split}
\langle \pi^0 e^+|\mathcal{O}_{\Gamma \Gamma'}|p\rangle =-\langle \pi^0 e^+|\mathcal{O}_{\tilde{\Gamma}\tilde{\Gamma}'}|p\rangle
\end{split}\end{align}
\noindent where $\tilde{R}=L$ and $\tilde{L}=R$. Using this, the decay width is given by \cite{DM, Cooney}:
\begin{align}\begin{split}
\Gamma(p\to \pi^0+e^+)=&\frac{m_p}{32\pi}\left(1-\left(\frac{m_\pi}{m_p}\right)^2\right)^2\times\\&\left|\sum_i C^i W^i_0(p\to \pi^0+e^+)\right|^2
\end{split}\end{align}
\noindent where the $W_0^i$ are the form factors associated to the operators $Q^{(i)}$ (i.e. $W_0^i=\langle \pi^0,e^+|Q^{(i)}|p\rangle$) and the $C^i$ are the Wilson coefficients coming from the renormalization of gauge couplings. These coefficients are given by running the coupling constants from mass scale from $\mu_0$ down to $\mu$ \cite{AbbottWise,DM}:
\begin{align}\begin{split}
C_1(\mu)&=\left[\frac{\alpha_3(\mu)}{\alpha_3(\mu_0)}\right]^{\frac{-2}{b_3}}\left[\frac{\alpha_2(\mu)}{\alpha_2(\mu_0)}\right]^{\frac{-9}{4b_2}}\left[\frac{\alpha_1(\mu)}{\alpha_1(\mu_0)}\right]^{\frac{-11}{20 b_1}}C_1^{\mu_0}\\
C_2(\mu)&=\left[\frac{\alpha_3(\mu)}{\alpha_3(\mu_0)}\right]^{\frac{-2}{b_3}}\left[\frac{\alpha_2(\mu)}{\alpha_2(\mu_0)}\right]^{\frac{-9}{4b_2}}\left[\frac{\alpha_1(\mu)}{\alpha_1(\mu_0)}\right]^{\frac{-23}{20 b_1}}C_2^{\mu_0}\\
C_3(\mu)&=\left[\frac{\alpha_3(\mu)}{\alpha_3(\mu_0)}\right]^{\frac{-2}{b_3}}\left[\frac{\alpha_2(\mu)}{\alpha_2(\mu_0)}\right]^{\frac{-15}{2b_2}}\left[\frac{\alpha_1(\mu)}{\alpha_1(\mu_0)}\right]^{\frac{-1}{10 b_1}}C_3^{\mu_0}\\
C_4(\mu)&=\left[\frac{\alpha_3(\mu)}{\alpha_3(\mu_0)}\right]^{\frac{-2}{b_3}}\left[\frac{\alpha_1(\mu)}{\alpha_1(\mu_0)}\right]^{\frac{-13}{5 b_1}}C_4^{\mu_0}\\
\end{split}\end{align}
Here $C_i^{\mu_0}$ is the coupling constant in the effective Lagrangian at the scale $\mu_0$ and $b_i$ is from the $\beta_i$ function. If the $\beta$ function changes with energy scale, such as when there are intermediate symmetry stages or scalar bosons with low mass, then one must run the Wilson coefficients several times starting from the unification scale (where the denominator starts with $\alpha_U$)\cite{AbbottWise,DM}.

\begingroup
\centering
\includegraphics[width=75 mm,keepaspectratio]{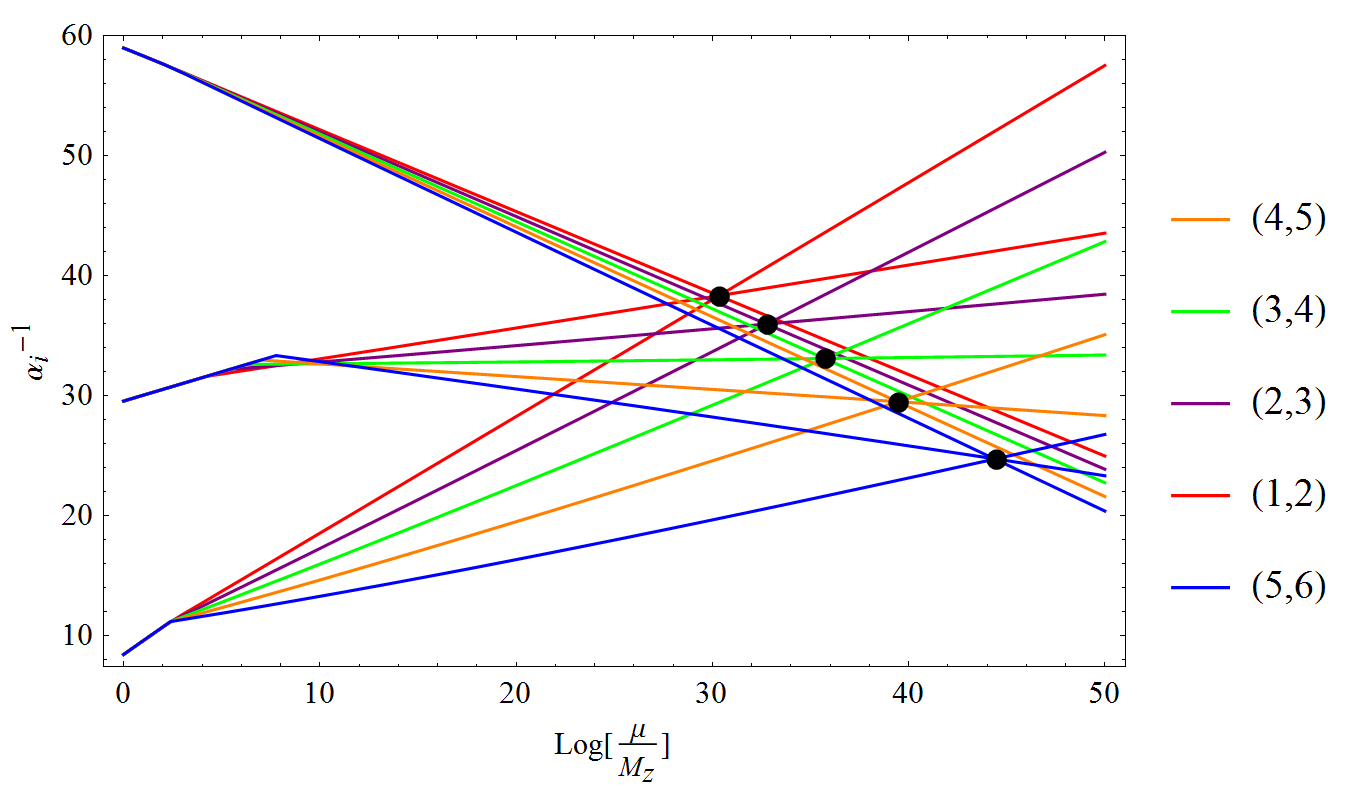}
\captionof{figure}{2-loop unification of the extended Babu-Mohapatra models. These are labeled as $(N_\Delta,N_\omega)$, which correspond to the number of color sextets and weak scalar triplets. }%normalized to the mass of the Z boson
\endgroup
\hspace{3mm}
For this model only the $\mathcal{O}_{LR}$ and $\mathcal{O}_{RL}$ operators are in the effective Lagrangian after integrating out the heavy gauge bosons. Using 2 GeV as the energy scale at which the proton decays (the standard in Lattice QCD calculations), we find that for this model:
\begin{align}\begin{split}
C_1=A_1 \frac{g_U^2}{M_U^2}=11.58\times \frac{g_U^2}{M_U^2}\\
C_2=A_2 \frac{g_U^2}{M_U^2}=12.97\times \frac{g_U^2}{M_U^2}
\end{split}\end{align}
Now that the Wilson coefficients have been calculated, all that is required to determine the proton lifetime are the QCD form factors. 

As explained before, these form factors must be computed using lattice QCD simulations. We will use the results from \cite{Aoki} which calculates the value:
\begin{align}
W_0^1=\langle \pi^0|\mathcal{O}_{RL}|p\rangle=-0.103(23)(34)
\end{align}
%\begin{flushleft}
%\begin{tabular}{|c|l|l|}
%  \hline
%  \textbf{Author}&  $\langle \pi^0|\mathcal{O}^{LL}|p\rangle$ \small{GeV} & $\langle \pi^0|\mathcal{O}^{RL}|p\rangle$ \small{GeV}\\\hline
%  Cooney \cite{Cooney,CooneyPub} & 0.147(33) & -0.139(31)\\ \cline{2-3}
%  & 0.079(55) & -0.125(77)\\
%  \hline 
%  Aoki et al. \cite{Aoki}& 0.133(29)(28)&-0.103(23)(34)\\
%  \hline\hline
%	\textbf{Average} & 0.120(25)&-0.122(26)\\
%\hline
%\end{tabular}
%\end{flushleft}
%\captionof{table}{Comparison of values for the hadronic form factors computed at renormalization scale $\mu=2$ GeV. In \cite{Cooney} the author computes the factors by two different methods. In \cite{18} they derive the same low energy QCD constants: $\alpha,\beta$. The average and average error are computed using basic error propagation techniques.}
%\hspace{3mm}\\
where the first error is statistical and the second is systematic. We now have the final formula for the proton decay width:
\begin{align}\begin{split}
\Gamma(p\to\pi^0\text{+}e^+)&=\frac{m_p}{32\pi}(A_1-A_2)^2\left(\frac{g_U^2W_0^1}{M_U^2}\right)^2\\&\times\left(1-\left(\frac{m_\pi}{m_p}\right)^2\right)^2
\end{split}\end{align}
\noindent Using the values given above for the QCD factor and the renormalized Wilson coefficients, we find:
\begin{align}\begin{split}
\tau(p\to\pi^0+e^+)\simeq &1.1\times 10^{33^{+0.3}_{-0.4}} \text{ yrs.} 
\end{split}\end{align}
\noindent where the error comes from the QCD form factor. In addition to this, we will also have errors coming from threshold effects. A short calculation shows that threshold effects lead to uncertainty in the unification scale:
\begin{align}\begin{split}
\Delta \log \frac{M_Z}{M_U}=0.374 \eta_{126}+0.011 \eta_{10}\\
+0.006 \eta_{45}+0.037 \eta_{\nu_R}
\end{split}\end{align}
where $\eta_i=\log\frac{M_i}{M_U}$. Following the prescription of \cite{ParidaPatra,MohapatraParida,LeeMohapatra,Hall} we allow the $M_i/M_U$ to vary between $10$ and $10^{-1}$ to give an upper bound on the estimate:
\begin{align}
M_U/M_U^{0}=10^{\pm0.43}
\end{align}
where $M_U^{0}$ is the central value of the unification scale. This leads to the proton lifetime of the order:
\begin{align}
\tau(p\to\pi^0+e^+)\simeq &1.1\times 10^{33^{+0.3}_{-0.4}\pm 1.72} \text{ yrs.} 
\end{align}
Taking into account the uncertainty from the lattice QCD calculation and threshold effects, we can see that the lifetime of the proton is outside the lower limit set by the Super Kamiokande experiment ($1.4\times10^{34}$ years \cite{Kamiokande,KamiokandePrelim}), but easily within reach of future experiments.

\section{Extension of the Babu-Mohapatra Model}

\indent It is interesting to note that the proton lifetime of this model could be lengthened by extending the Babu-Mohapatra model. This can easily be achieved by adding more non-standard model scalars ($\Delta_{u^cd^c}$, $\omega$) whose numbers we will denote $(N_\Delta,N_\omega)$ for the number of sextets and triplets respectively. This causes the unification scale to increase while coupling at unification decreases. Some values are plotted in Figure 2 with corresponding unification values in Table 1.

\begin{center}
\begin{tabular}{|c|c|c|}
  \hline
  ($N_\Delta,N_\omega$)&  $ m_\omega$ (TeV)&$\tau(p\to \pi^0+e^+) \text{ yrs.}$\\\hline
(1,2)&6.2&$1.1\times10^{33^{+0.3}_{-0.4}\pm 1.72}$\\
\hline
(2,3)&21&$3.8\times10^{37^{+0.3}_{-0.4}\pm 3.21}$\\
\hline
(3,4)&48&$9.8\times10^{44^{+0.3}_{-0.4}\pm 4.72}$\\
\hline
(4,5)&100& $5.4\times10^{43^{+0.3}_{-0.4}\pm 6.22}$\\
\hline
(5,6)&220&$3.8\times10^{53^{+0.3}_{-0.4}\pm 7.73}$\\
\hline
\end{tabular}\end{center}
\captionof{table}{Comparison of extensions to the Babu-Mohapatra model by adding additional color sextets ($\Delta$) and weak triplets ($\omega$). In these models $m_\Delta=1$ TeV.}
\hspace{3mm}

For these models we have fixed the sextet scalar mass $m_\Delta=1$ TeV in order to showcase the extensions with new TeV level physics. It seems that in general, full two loop unification with low TeV scalar masses can only be achieved if there is 1 more triplet than sextet (although it is believed that all extensions will unify with higher scalar masses). These extended Babu-Mohapatra models provide a large parameter space that can be compared with experimental bounds on dark matter from experiments such as LUX and the LHC while maintaining a sufficiently long proton lifetime. Further study of these models is required. 

\section{\textbf{Conclusion}}

We have demonstrated in this paper, that the non-SUSY SO(10) GUT model put forth by Babu and Mohapatra in \cite{BM} unifies to two loop order at $M_U=1.42\times 10^{15\pm0.43}$ GeV with scalars at the TeV scale \cite{BM}. We have presented the calculations to show that this model predicts the proton lifetime to be $1.1\times 10^{33^{+0.3}_{-0.4}\pm 1.72}$ yrs, which places the proton lifetime above of the Super-Kamiokande limit \cite{Kamiokande}. In addition we have also proposed a class of extensions of the Babu-Mohapatra model which incorporate additional scalar fields and have the additional merits of a longer proton lifetime and dark matter candidates with higher mass. These models have been shown to meet the minimum requirement for proton lifetime and therefore should be further investigated for their rich low energy phenomenology. Because of the inclusion of low mass scalar sextets and triplets, this model will potentially produce many physical effects which will be observable in the next generation of experiments. In addition to the prediction of particle discoveries at the LHC, this model is also of great interest due to its predictions of neutron-antineutron oscillation, new mechanism for scalar mediated baryogenesis, and dark matter candidates \cite{BM}. 

\section{\textbf{Acknowledgements}}

This research was supported in part by the U.S. Department of Energy under grant DE-FG02-96ER40959. The author would like thank Rabindra Mohapatra for guidance, assistance in writing this paper, and proposing this project. The author would also like to thank David Shih, Matt Buckley, and Angelo Moneaux for enlightening conversations and Pouya Asadi for enlightening discussion and comments. 

%\bibliographystyle{iopart-num}
%\bibliography{2loop}{}
%\end{multicols}

\end{document}